\documentclass[10pt,conference,a4paper]{IEEEtran}

\usepackage[utf8]{inputenc}

\usepackage{blindtext, graphicx}
\usepackage[utf8]{inputenc}
\usepackage[english]{babel}
\usepackage{csquotes}
\usepackage{wrapfig}
\usepackage{float}
\usepackage{verbatim}
\usepackage{subcaption}
\usepackage[font=footnotesize,labelfont=bf]{caption}
\usepackage{import}
\usepackage{dirtytalk}
\usepackage{subfiles}
\usepackage{array}
\usepackage{multirow}

\usepackage{lscape}
\usepackage{algorithm}
\usepackage{algorithmic}
\usepackage{rotating}
\usepackage[table,xcdraw]{xcolor}

\usepackage{longtable} 
\usepackage{array} 
\usepackage{caption} 
\usepackage{graphicx} 
\usepackage{xcolor, colortbl} 
\usepackage{booktabs} 
\usepackage{tabularx} 
\usepackage{amsmath}

\usepackage{times}

\DeclareGraphicsExtensions{.png,.eps,.ps,.pdf}

\usepackage{url}
\usepackage{hyperref}

\usepackage[most]{tcolorbox}

\usepackage{multirow}
\usepackage{tabularx}
\usepackage{makecell}
\usepackage{svg}

\usepackage{pifont}
\definecolor{ao}{rgb}{0.0, 0.5, 0.0}
\definecolor{amber}{rgb}{1.0, 0.49, 0.0}

\let\labelindent\relax
\usepackage{enumitem}

\usepackage{lscape} 

\graphicspath{{Figures/}}

\usepackage{tikz} 
\newcommand{\orcidicon}{\includegraphics[width=0.32cm]{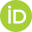}}
\foreach \x in {A, ..., Z}{%
\expandafter\xdef\csname orcid\x\endcsname{\noexpand\href{https://orcid.org/\csname orcidauthor\x\endcsname}{\noexpand\orcidicon}}
}

\begin{document}

\title{Hybrid Tabletop Exercise (TTX) based on a Mathematical Simulation-based Model for the Maritime Sector}

\author{\IEEEauthorblockN{
\orcidA{}Diego Cabuya-Padilla$^{1}$,
\orcidB{}Daniel D\'iaz-L\'opez$^{2,3}$,
\orcidC{}Carlos Castaneda-Marroqu\'in$^{1}$
}

\IEEEauthorblockA{$^1$ ``Cyberspace, Technology and Innovation'' research group, Escuela Superior de Guerra ``General Rafael Reyes Prieto'', \\ Cra. 11 \# 102-50, Bogot\'a, Colombia}
\IEEEauthorblockA{$^2$Department of Information and Communications Engineering, University of Murcia, 30100, Murcia, Spain\\
danielorlando.diaz@um.es}
\IEEEauthorblockA{$^3$School of Engineering, Science and Technology, Universidad del Rosario, Bogot\'a, Colombia\\
danielo.diaz@urosario.edu.co}
}


\maketitle

\begin{abstract}
As cyber threats grow in complexity and scale, many security incidents remain poorly managed due to the lack of proper training among C-level executives. Thus, there is a need for targeted cybersecurity education to enhance executive decision-making and crisis response. Traditional training methods, such as cyber wargames and Tabletop Exercises (TTX), aim to develop abilities to face critical incidents, however, they often lack the interactive and dynamic elements required to prepare individuals for real-world cyber incidents. This paper presents a novel approach to cybersecurity and cyberdefense education through the design of a specialized hybrid TTX for the maritime domain, which uses a framework to model mathematically how a cyberattack spreads along multiple nodes and impacts infrastructure. Our proposal was validated through exercises in Argentina and the United States, demonstrating a positive impact in developing the comprehension and projection levels of Cyber Situational Awareness (CSA), and reinforcing governance. Documentation about the Hybrid TTX, scenario, datasets and implementation of the SERDUX-MARCIM model, is available at the project repository
at \url{https://github.com/diegocabuya/SERDUX-MARCIM}

\end{abstract}

\begin{IEEEkeywords}
Cybersecurity, Cyberdefense, Maritime, Serious Games, Tabletop Exercises (TTX), Cyber Situational Awareness (CSA), Education in cybersecurity
\end{IEEEkeywords}

{\bf Contribution type:}  {\it  Training and educational innovation}

\section{Introduction}\label{intro}

Tabletop Exercises (TTX) are structured discussion-based activities designed to train teams in responding to cybersecurity incidents and crisis situations~\cite{nist2006guide}. Unlike hands-on technical training, TTX focuses on decision-making, communication, and coordination under simulated emergency conditions. These exercises provide a controlled environment where participants assume organizational roles and collaboratively address evolving cyber threats through scenario-driven discussions~\cite{ENISA2009}. TTX's are widely used in cybersecurity training and risk management, as they help organizations evaluate their response capabilities and identify gaps in policies and procedures. Their adaptability and cost-effectiveness make them a valuable tool for improving resilience against cyberattacks while fostering a culture of proactive security awareness~\cite{ISO22398}.

In addition, Cyber Situational Awareness (CSA) is a fundamental element, as it enables organizations to understand and anticipate cyber threats by processing and analyzing relevant information from various sources~\cite{Franke2014}. CSA is based on the broader concept of situational awareness, which consists of 3 levels: perception, comprehension, and projection~\cite{Endsley1995}. The perception level involves identifying key cyber elements, such as anomalies in network traffic or security alerts. The comprehension level integrates and analyzes this information to determine its significance, identifying potential threats or vulnerabilities. Finally, the projection level allows decision makers to anticipate the evolution of cyber incidents, assessing their potential impact, and determining proactive response strategies. These levels collectively improve an organization's ability to detect, assess, and mitigate cyber risks, forming a crucial foundation for proactive cybersecurity.

TTX establishes a structured and controlled scenario that allows participants to analyze decision-making processes and response strategies to cyber incidents~\cite{angafor2020}. However, traditional TTX methodologies face significant challenges, such as limited real-time data integration, lack of dynamic adversarial behavior modeling, and difficulties in quantitatively assessing decision-making~\cite{vsvabensky2024}. Thus, to address the challenges identified in traditional TTX methodologies, this paper proposes an evolution towards a hybrid TTX approach that integrates serious games and simulation-based techniques. The main contributions of this paper are summarized as follows:


\begin{itemize}
    \item Design of a novel Hybrid TTX Methodology that integrates the traditional TTX approach with a mathematical simulation-based model (SERDUX-MARCIM).
    \item Application of the proposed hybrid TTX methodology in the maritime sector, strengthening CSA levels among strategic, including military personnel, maritime authorities, and policymakers. 
    \item Evaluation of our proposal in 2 real-World Scenarios, one conducted in Argentina, and the other in United States. 
\end{itemize}

This paper is structured as follows: Section~\ref{background} introduces the SERDUX-MARCIM model. Section~\ref{method} describes our proposal. Section~\ref{experiments} and Section~\ref{sec:prospective_analysis} details the execution of two TTX exercises, highlighting participant insights on CSA. Section~\ref{conclusions} presents the conclusions and outlines future work.

\section{State of the art}\label{sota}

A bibliometric review reveals a lack of specialized tools to support the strategic adoption of response protocols in maritime cybersecurity and cyberdefense. Nonetheless, some studies offer valuable insights for the proposal.

Damodaran and Wagner~\cite{Damodaran2020} highlight simulation's role in cyberdefense through game-theoretic models and network recovery strategies, stressing its value for cyber situational awareness (CSA). Similarly, Bodeau et al.~\cite{Bodeau2018} and Katsantonis~\cite{Katsantonis2019} present wargaming frameworks linking cybersecurity training to risk management. Valente and Reith~\cite{Valente2024} advocate serious gaming over traditional training in military settings, emphasizing the inclusion of IT and OT systems to improve awareness in critical infrastructure. Onduto~\cite{Onduto2021} supports gamification's impact on behavior but warns of its experimental maturity and limited assessment of long-term learning.

In summary, simulation-based methods have enhanced cybersecurity education, but critical gaps persist in the maritime sector. Current tools often fail to capture the complexity of interconnected maritime systems, signaling the need for novel, simulation-driven methodologies to reinforce strategic preparedness and structured cyber response.
\section{Background}\label{background}

As part of the MARCIM research project~\cite{Cabuya-Padilla_Castaneda-Marroquin_2024}, SERDUX-MARCIM~\cite{Cabuya-Padilla2025} was developed as a mathematical and simulation model to support experimenters in formulating and testing working hypotheses, test courses of action, and identifying potential maritime cybersecurity and cyberdefense scenarios. 

SERDUX-MARCIM combines three techniques: compartmental epidemiological models, system dynamics, and agent-based modeling. This combination allows for a more comprehensive understanding of how cyberattacks propagate within a maritime ecosystem, simulating the attack as a dynamic process. By adapting epidemiological models such as SIR (Susceptible, Infected, Recovered) and SEIR (Susceptible, Exposed, Infected, Recovered) to the context of cyberattack propagation, this model offers a way to assess risk and predict the behavior of such attacks as they unfold, making it relevant for both operational validation and strategic awareness.

Finally, SERDUX-MARCIM enhances the three strategic levels of CSA, i.e. perception, comprehension, and projection, by offering a framework that guides decision-making during cyberattacks on maritime infrastructure, improving response coordination and resilience in maritime cybersecurity.

\section{Hybrid Tabletop Exercise Methodology for Maritime Sector}\label{method}

A Hybrid TTX methodology is proposed in this section integrating the traditional TTX approach~\cite{grance2006guide} with the SERDUX-MARCIM model~\cite{Cabuya-Padilla2025}, introduced in Section~\ref{background}. By incorporating serious gaming techniques, this approach allows participants to interact with the computational model, gaining deeper insights into its implications within a controlled yet realistic decision-making environment.

In this context, the hybrid TTX is designed to strengthen maritime cybersecurity and cyberdefense by providing a high-level perspective. A key emphasis is placed on CSA as a fundamental component of informed decision-making across strategic, tactical, and operational levels.

\subsection{Scenario}

Participants worked with the 2017 NotPetya cyberattack against Maersk~\cite{Abbatemarco_2024_Maersk,Use_Case_Maersk_HBR, Sawang2023} as a reference case to implement a case-based learning strategy aimed at developing CSA levels. This scenario was selected due to its relevance as one of the most disruptive cyber incidents in the maritime sector, exposing critical vulnerabilities in maritime infrastructures ~\cite{Cabuya-Padilla2025}. By using this cyberattack, the exercises allowed participants to reflect on the complexities of cyber threats in the maritime domain and the importance of comprehensive risk assessments, as well as the design of effective strategies.

\begin{figure}[ht!]
    \centering
    \includegraphics[width=\columnwidth]
    {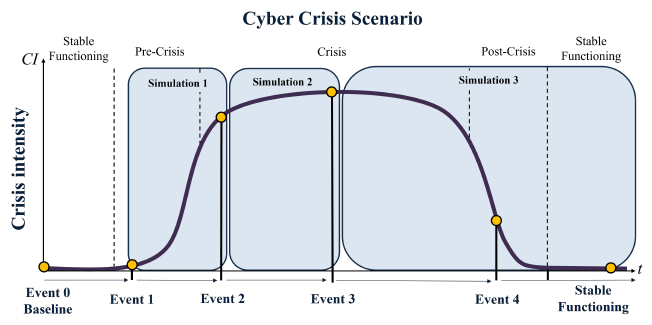}
    \caption{Simulations and events defined for the application of SERDUX-MARCIM framework in the Maersk case~\cite{Cabuya-Padilla2025}}
    \label{fig:Cyber_Risk_Methodology}
\end{figure}

Within the SERDUX-MARCIM framework, an Event $i$ represents a key milestone in the progression of a cyber crisis scenario and it belongs to any of the 3 standard crisis phases~\cite{Sawang2023}: pre-crisis, crisis, and post-crisis. Thus, for the application of the SERDUX-MARCIM framework in a cyberdefense scenario, we must define the key variables indicated in Table~\ref{tab:validation_variables}. To apply SERDUX-MARCIM in the Maersk case, we consider the execution of 3 simulations, i.e. $s=3$, 4 key events along the scenario, i.e. $i=4$, and 3 variables to monitor, i.e. $pm=3$. Specifically, the variables to monitor in this scenario are i) network situation, ii) level of service availability, and iii) overall cyber risk posture. This setup is illustrated in Figure~\ref{fig:Cyber_Risk_Methodology}. This structured approach allows participants to systematically evaluate how cyber threats evolve and how decision-making dynamics shift throughout a crisis.

\subsection{Methodology}

\begin{figure}[H]
    \centering
    \includegraphics[width=\columnwidth]
    {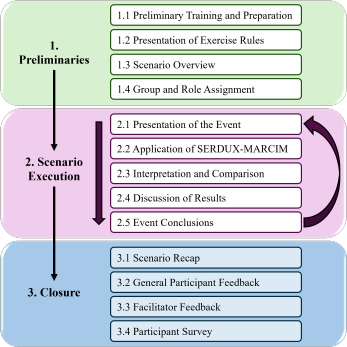}
    \caption{Methodology proposed for the hybrid TTX}
    \label{fig:TTX_Methodology}
\end{figure}

The proposed Hybrid TTX methodology consists of seven steps, incorporating a structured decision-making process supported by the SERDUX-MARCIM model. This hybrid approach enhances CSA at a strategic level by engaging participants in simulated cyber crisis scenarios. While the methodology contributes to the development of all CSA levels, it primarily strengthens the comprehension and projection skills of participants, allowing them to interpret complex situations and anticipate potential outcomes. The proposed methodology, along with its phases and steps, is depicted in Figure~\ref{fig:TTX_Methodology}.

The preliminary phase is designed to prepare participants and establish the conditions for the successful execution of the exercise. Its main purpose is to standardize participants' knowledge, define the rules and structure of the activity, and ensure that everyone clearly understands the scenario and their assigned roles. This phase lays the groundwork for meaningful discussions and informed decision-making during the exercise.

The scenario execution phase consists of the dynamic development of the cybersecurity scenario through sequential events. Each event follows a cyclical process, repeating as many times as the scenario establishes. Each cycle is composed of five steps describe below:
\begin{itemize}
    \item \textbf{Presentation event $i$}: Event $i$ is presented to participants, providing relevant details about it: organizational context, key actors involved, and any critical information necessary to understand the situation.
    \item \textbf{Application of the SERDUX-MARCIM}: The event $i$ is integrated in the SERDUX-MARCIM model, which provides a structured cyber risk analysis. This allows participants to dynamically assess the evolution of the attack in terms of impact on maritime infrastructure.
    \item \textbf{Interpretation and comparison}: Participants analyze the outputs generated by the SERDUX-MARCIM model, focusing on understanding the current situation (comprehension) and anticipating potential future developments (projection) of the cyberattack.
    \item \textbf{Discussion of results}: Facilitators use carefully designed guiding questions to promote a structured discussion aimed at strengthening participants' CSA, with a particular focus on developing their Projection capabilities. 
    \begin{itemize}
        \item First, participants are invited to analyze the current situation of the event, reflecting on the implications for the maritime sector, identifying affected services, and assessing the role of internal and external actors involved. 
    \end{itemize}
    \begin{itemize}
        \item Second, the discussion shifts toward the formulation of possible courses of action. Participants evaluate response alternatives based on the current scenario dynamics and the projected impacts of their decisions as estimated by the model outputs. 
    \end{itemize}
    \begin{itemize}
        \item Finally, participants are guided to deliberate and agree on the most favorable course of action. 
    \end{itemize}
    \item \textbf{Event conclusions}: The group leader presents the most relevant insights and conclusions based on the guiding questions. These conclusions summarize key takeaways, lessons learned, and potential improvements in cybersecurity and cyberdefense strategies.
\end{itemize}

The closure phase is in charge of consolidating the key lessons learned, evaluating the exercise outcomes, and gathering participant feedback to strengthen future training activities. This phase aims to promote collective reflection on the decisions made, the strategies discussed, and the overall development of CSA throughout the exercise.

\section{Experiments}\label{experiments}

This section describes the experiments conducted in Buenos Aires, Argentina, and Washington D.C, United States, where the hybrid TTX proposed in Section~\ref{method} was applied to assess its effectiveness in fostering CSA at the strategic level. 

\subsection{Participants}
The exercises involved 36 high-level international stakeholders with strategic responsibilities in maritime security. Participants’ roles included military personnel, maritime authorities, policymakers, deputy ministers, ambassadors, diplomats, academics, and cybersecurity specialists. The selection criteria prioritized individuals with extensive experience in strategic decision-making and crisis management. Notably, given the strategic focus of the exercise, participants were not undergraduate or graduate students, but professionals actively engaged in roles tied to maritime power. Gathering such a distinguished and diverse group, spanning 16 countries, mainly from the Americas, is particularly significant; the sample size is substantial considering the highly specific profile required, which makes this level of participation exceptionally rare in conventional training environments.

\subsection{Exercises Description}

\subsubsection{Exercise 1} It was conducted in Buenos Aires, Argentina, on February 23, 2024, during the \textit{SALMA Dialogue: For a More Efficient Maritime Security in the Central and South Atlantic}. The variables considered for the exercise are listed in Table~\ref{tab:validation_variables}. The methodology involved presenting SERDUX-MARCIM model simulations applied to the Maersk cyberattack within a structured TTX format. This approach integrated plenary sessions for scenario analysis and small group discussions (3–4 participants), allowing in-depth examination of key cybersecurity challenges. Following the exercise, participants completed an individual survey, which will be described in Section~\ref{sec:DataCollection}, to assess their experience and provide feedback on the model's applicability.

\begin{table}[H]
    \centering
    \begin{tabular}{c c c c} \hline 
         \textbf{Variable}& \textbf{Symbol} & \textbf{Exercise 1} & \textbf{Exercise 2}\\ \hline 
         Number of Simulation& $s$ & 3 & 3\\ 
         Number of events to consider& $i_t$ & 5 & 5\\ 
         Perspectives to monitor per event& $pm$ & 3 & 3\\  
         Number of Participants& $np$ & 10 & 26\\ 
         Number of Observers& $no$ & 15 & 4\\  
         Group Size& $gs$ & 3-4 & 8-12\\ \hline 
    \end{tabular}
    \caption{Variables defined per exercise}
    \label{tab:validation_variables}
\end{table}

\subsubsection{Exercise 2} It took place in Washington, D.C., United States, from May 16 to 17, 2024, as part of the \textit{Advanced Course on Maritime Security Policies}, organized by the William J. Perry Center at the National Defense University, U.S. The variables considered for the exercise are related in Table~\ref{tab:validation_variables}, and the exercise followed a similar methodology to Exercise 1.

\subsection{Data Collection Method}\label{sec:DataCollection}

According to the methodology outlined in Section~\ref{method}, the final step of the \textit{Closure} phase involves the application of a \textit{Participant Survey}. Accordingly, two surveys were conducted, comprising 18 variables [X1–X18] as detailed in Table~\ref{tab:variables_description_t2}. These variables were selected to ensure a comprehensive evaluation of the exercise, encompassing aspects related to the participants, the exercise execution, and the scenario design.

\begin{itemize}
    \item \textbf{Survey 1:} TTX assessment, structured into three sections:
    \begin{enumerate}
        \item Participants' previous experience with TTX or simulations, particularly in the maritime domain, as well as their training in cybersecurity or cyberdefense. [X1-X4]
        \item General evaluation of the exercise. [X5-X8]
        \item Scenario evaluation. [X9-X13]
    \end{enumerate}
    \item \textbf{Survey 2:} SERDUX-MARCIM model overall accuracy. [X14-X18]
\end{itemize}


\begin{table*}[t]
\centering
\renewcommand{\arraystretch}{1.3} 
\setlength{\tabcolsep}{4pt} 
\begin{tabularx}{\textwidth}{|c|c|X|X|X|} 
\toprule
\textbf{Variable} & \textbf{Range} & \textbf{Name} & \textbf{Variable Definition} & \textbf{Guiding Question for Data Collection} \\
\midrule
Y  & 0 to 5 & Overall Satisfaction & Participants' overall perceived satisfaction with the TTX exercise, considering all aspects of the experience. & Evaluate your overall satisfaction level with the TTX. \\
\hline
X1 & 0 or 1 & Participation in TTX & Participant's prior experience in crisis simulation exercises or Tabletop Exercises (TTX), which may influence their perception of the current exercise. & Have you ever participated in a Tabletop Exercise (TTX) or a crisis simulation exercise? \\
\hline
X2 & 0 or 1  & Participation in CS\&CD Simulations & Participant’s background in cybersecurity and cyberdefense simulations, which may affect their familiarity with TTX concepts and dynamics. & Have you ever participated in a cybersecurity or cyberdefense simulation exercise? \\
\hline
X3 & 0 or 1  & Participation in Maritime CS\&CD Simulations & Participant’s prior experience in simulations specifically focused on cybersecurity and cyberdefense in the maritime domain. & Have you ever participated in a cybersecurity or cyberdefense simulation exercise in the maritime domain? \\
\hline
X4 & 0 or 1  & Cybersecurity and Cyberdefense Training & Academic background, training, or informal education received by the participant in cybersecurity and cyberdefense. & Have you received any formal or informal training in cybersecurity or cyberdefense? \\
\hline
X5 & 0 to 5  & Facilitator Effectiveness & Clarity and effectiveness of the instructions, guidance, and assistance provided by facilitators during the TTX. & How clear were the instructions and assistance provided by the facilitators? \\
\hline
X6 & 0 to 5 & Time Management & Assessment of time management during the exercise, including task distribution and the adequacy of time allocated to each phase of the TTX. & How well was time managed during the TTX? \\
\hline
X7 & 0 to 5 & Resources & Quality and adequacy of the materials provided (e.g., game instruction booklet) to facilitate understanding and execution of the exercise. & Did the booklet (game instructions) contain sufficient information for conducting the TTX? \\
\hline
X8 & 0 to 5 & Scenario Relevance & The relevance of the TTX to maritime security awareness and its applicability to participants' operational context. & How relevant was this maritime cybersecurity TTX for raising awareness of maritime security? \\
\hline
X9 & 0 to 5 & Scenario Information Relevance & Value and usefulness of the general scenario information in understanding events and their implications. & How relevant was the general scenario information for understanding the situation and the implications of each event? \\
\hline
X10 & 0 to 5 & Scenario Understanding & Participants’ level of comprehension of the presented scenario, considering the information provided for each event of the exercise. & How would you rate your overall understanding of the SCENARIO presented, considering the information provided per event? \\
\hline
X11 & 0 to 5 & Scenario Complexity & Perceived difficulty level of the scenario in terms of exercise structure, involved variables, and presented challenges. & How would you rate the overall complexity of the scenario? \\
\hline
X12 & 0 to 5 & Scenario Detail Level & Accuracy and specificity of the simulated scenario elements, including the fidelity of event and situation representation. & How would you assess the level of detail in the simulated scenario? \\
\hline
X13 & 0 to 5 & Application of knowledge and skills & Applicability of the knowledge and skills acquired during the exercise to the participant’s professional responsibilities in maritime cybersecurity. & How applicable are the skills and knowledge acquired in this maritime cybersecurity exercise to your professional responsibilities? \\
\hline
X14 & 0 to 5 & SERDUX-MARCIM Information Relevance & Degree to which the information provided by the SERDUX-MARCIM model was useful and pertinent for understanding the scenario and its implications. & How relevant do you believe the information related to SERDUX-MARCIM was to understand the situation and implications in every event? \\
\hline
X15 & 0 to 5 & SERDUX-MARCIM Information Accuracy & Degree of precision and correctness of the information and visual outputs generated by the SERDUX-MARCIM model. & Considering the scenario presented, how precise do you believe the information and graphics related to the SERDUX-MARCIM were? \\
\hline
X16 & 0 to 5 & SERDUX-MARCIM Need for Similar Models Across Organizations & Perceived importance of implementing simulation models like SERDUX-MARCIM to support decision-making during cyber crisis situations. & How relevant do you believe that implementing cyber threat simulation models (like SERDUX-MARCIM) is for decision-making in a crisis situation? \\
\hline
X17 & 0 to 5 & SERDUX-MARCIM Integration & Perceived level of difficulty in integrating simulation models like SERDUX-MARCIM into current maritime cybersecurity protocols and organizational frameworks. & How complex do you perceive integrating cyber threat simulation models (like SERDUX-MARCIM) into existing maritime cybersecurity protocols and frameworks within your organization? \\
\hline
X18 & 0 to 5 & SERDUX-MARCIM Expected Applicability of the model & Extent to which the SERDUX-MARCIM model's output behavior met the required accuracy to achieve the exercise objective. & Was the output behavior of the SERDUX-MARCIM model sufficiently accurate to meet the exercise objective? \\
\hline
X19 & 0 to 5 & General Observations & Comments or observations related to the exercise development. & N/A \\
\bottomrule
\end{tabularx}
\caption{Definition of Variables and Guiding Questions (Variable Type: numerical)}
\label{tab:variables_description_t2}
\end{table*}

Data collected through these surveys was analyzed to identify statistical associations and performance patterns, providing insights into participants' perceptions regarding the effectiveness of the exercise, the applicability of the SERDUX-MARCIM model in real-world scenarios, and the appropriation of CSA skills (Perception, Comprehension, Projection) at the strategic level. To ensure data protection, sensitive participant information, such as names and affiliations, was omitted.

\subsection{Survey Results}\label{sec:surveys}

\subsubsection{Survey 1 - TTX Assessment}\label{sec:survey1}

\paragraph{Section 1 - Participants’ Prior Experience and Cybersecurity or Cyberdefense Training}

\begin{figure}[H]
    \centering
    \includegraphics[width=\columnwidth]
    {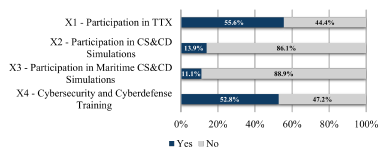}
    \caption{Survey 1 - Participants' Prior Experience and Training [0\% - 100\%] }
    \label{fig:Survey1_A}
\end{figure}

Findings (Figure~\ref{fig:Survey1_A}) indicate that only 55.6\% of the participants had previously participated in a TTX or a crisis simulation exercise ($X1$). Additionally, 13.9\% had taken part in cybersecurity and cyberdefense simulation exercises ($X2$), and 11.1\% had participated in maritime cybersecurity and cyberdefense simulation exercises ($X3$). Moreover, 52.8\% of the participants had received some formal or informal training in cybersecurity ($X4$).

The analysis highlights a significant lack of engagement in cybersecurity simulation exercises, particularly in the maritime sector. Furthermore, the results emphasize the significant need for enhanced training, especially at the strategic level, to strengthen maritime cybersecurity preparedness.

\paragraph{Section 2 - General Evaluation of the Exercise}

\begin{figure}[H]
    \centering
    \includegraphics[width=\columnwidth]
    {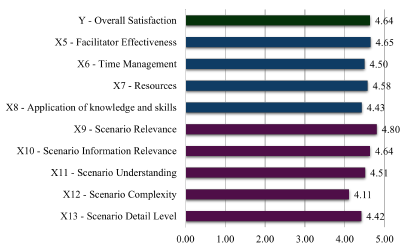}
    \caption{Survey 1 - General Evaluation - Scenario Evaluation [0 - 5]}
    \label{fig:Survey1_B}
\end{figure}

Figure~\ref{fig:Survey1_B} depicted that the exercises received a highly favorable rating, with an average score of 4.56/5, considering aspects such as facilitator effectiveness ($X5$), time management ($X6$), resources ($X7$), and overall participant satisfaction ($Y$). Participants expressed a high level of satisfaction and strong interest in replicating such exercises in their respective countries to improve cyber awareness and crisis management capabilities. Additionally, the exercises facilitated the internalization of fundamental concepts related to the typical evolution of cyber crises and the strategic responses required at each phase.

Despite the positive outcomes, participants suggested additional preparatory materials to optimize engagement, extending the duration of the exercise, and introducing a full-day cybersecurity and cyberdefense briefing beforehand.

\paragraph{Section 3 - Scenario Evaluation}

Figure~\ref{fig:Survey1_B} depicted that the scenario evaluation was also highly positive, with an average rating of 4.52/5. The assessed criteria included questions about different aspects of the proposed cyberdefense scenario: relevance ($X8$), information relevance ($X9$), understanding ($X10$), complexity ($X11$), level of detail ($X12$), and application of knowledge and skills ($X13$). Participants concluded that the scenario was highly relevant to the maritime sector, and the provided information was adequate.

However, due to the varying levels of prior cybersecurity knowledge among participants, some found the scenario somewhat complex, particularly regarding the technical aspects of the final events. As an area for improvement, participants recommended providing a clearer strategic interpretation of the tactical and operational situation during these events.

\subsubsection{Survey 2 - SERDUX-MARCIM Model Overall Accuracy}\label{sec:survey2}

\begin{figure}[H]
    \centering
    \includegraphics[width=\columnwidth]
    {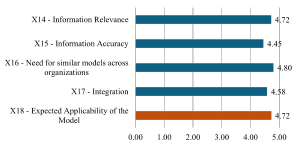}
    \caption{Survey 2 - SERDUX-MARCIM model overall accuracy [0 - 5]}
    \label{fig:Survey_2}
\end{figure}

The SERDUX-MARCIM model assessment (Figure~\ref{fig:Survey_2}) received highly positive feedback, with an average score of 4.67 out of 5 across key evaluation criteria: relevance ($X14$), accuracy ($X15$), the need for similar models in other organizations ($X16$), integration potential ($X17$), and expected applicability within the intended domain ($X18$).

These results underscore SERDUX-MARCIM's effectiveness as a tool for enhancing maritime cybersecurity awareness, particularly by improving the comprehension and projection dimensions of CSA. This improvement is supported from two perspectives. First, the high perceived value that participants assigned to the model-based outputs, especially regarding relevance ($X14$) and accuracy ($X15$), which aligns with the TTX objectives. Second, SERDUX-MARCIM's ability to provide participants with actionable information that enables the analysis of the current situation, the observation of variable trends over time, and the visualization of potential future behaviors. Specifically, the model facilitates the examination of three key elements: (i) the network situation, (ii) the level of service availability, and (iii) the overall cyber risk posture. This comprehensive perspective allowed participants to better understand the evolving dynamics of the simulated cyberattack and to anticipate its potential impacts on maritime operations.

Moreover, SERDUX-MARCIM demonstrates its potential as a decision-support tool in strategic responses to cyber threats, as reflected in the positive evaluations of the need for similar models across organizations ($X16$) and its integration capabilities ($X17$).

Finally, the results for the expected applicability of the model ($X18$) confirm that SERDUX-MARCIM achieves the required accuracy for its intended purpose and is well-suited for maritime cybersecurity and cyberdefense contexts. Participants particularly emphasized the model’s practical utility and the effectiveness of the tools and methods employed.

Additionally, regarding the general observations ($X19$) gathered through an open question, participants acknowledged that integrating models like SERDUX-MARCIM into their organizations could present certain challenges. However, there was significant interest in adopting similar initiatives. These findings reinforce the validity of the proposed framework and highlight its potential to strengthen cybersecurity and cyberdefense capabilities within critical maritime infrastructure.

\section{Prospective analysis}\label{sec:prospective_analysis}

The objective of this section is to provide a comprehensive prospective analysis aimed at identifying areas for improvement in the design and execution of future cybersecurity and cyberdefense training exercises, particularly within the maritime sector. This analysis is divided into two main sections: First, \textit{Multiple Linear Regression Model}, utilizes statistical modeling to predict the influence of key variables on participant outcomes. Finally, the \textit{Prospective Scenario Analysis} explores potential future scenarios—pessimistic, trend-based, and optimistic—to guide the refinement of future exercises based on the analysis conducted. Each section contributes to a deeper understanding of how to optimize the learning and engagement of participants in future exercises.

\subsection{Objective and Methodology}

To gain insights for improving future iterations of the proposed hybrid TTX, a mathematical analysis was conducted based on Survey 1 (Section~\ref{sec:survey1}). This analysis aimed to assess the influence of attributes related to i) participants, ii) TTX execution, and iii) presented scenario, on the overall satisfaction with the TTX and in the ability to apply the knowledge acquired during the TTX.

Two analytical approaches were applied: i) multiple linear regression modeling, and ii) prospective scenario analysis. The dependent variable for these analyses was ``Overall Satisfaction'' ($Y$), while independent variables included participant experience ($X1-X4$) and exercise perception ($X5-X13$).

\subsection{Multiple Linear Regression Model}

The multiple linear regression model developed for the collected data is expressed in Equation 1.

{\footnotesize
\begin{equation}
\begin{aligned}
Y = -0.0252 X1 + 0.8471 X2 - 0.7636 X3 - 0.0959X4 + 0.1139 X5\\
+ 0.2345 X6 + 0.4938 X7 + 0.1171 X8 + 0.0975 X9 - 0.0491 X10\\
+ 0.0090 X11 + 0.1979 X12 - 0.0631 X13 - 0.5722
\end{aligned}
\end{equation} \label{eq:regression}
}

The model demonstrates strong explanatory power ($R²=87.43\%$) and the observed high correlation among variables ($0.935$) confirms anticipated relationships. The adjusted R² is slightly lower ($80\%$) but still significant, indicating a robust and reliable predictive model. The standard deviation is relatively low ($0.26$) suggesting minimal dispersion. The variables with the greatest weight in the model include prior participation in cybersecurity and cyberdefense simulations ($X2$), prior participation in maritime cybersecurity and cyberdefense simulations ($X3$), time management($X6$), and resources($X7$).

\subsection{Prospective Scenario Analysis}

Based on the multiple linear regression model, three prospective scenarios were developed, as shown in Table~\ref{tab:scenarios}. These scenarios serve as reference frameworks, enabling future exercises to be compared against them in order to determine which scenario the observed results most closely align with.

\begin{table}[H]
    \centering
    \begin{tabular}{|c|c|c|c|} \hline 
         \textbf{Variable}&  \textbf{Pessimistic}&  \textbf{Trend-based}& \textbf{Optimistic}\\ \hline 
         Y& 2.8 & 3.8 & 4.8 \\ \hline 
         X1& 1 & 1 & 1\\ \hline 
         X2& 0 & 0 & 1\\ \hline 
         X3& 0 & 0 & 1\\ \hline 
         X4& 0 & 1 & 1\\ \hline 
         X5& 3 & 4 & 5\\ \hline 
         X6& 3 & 4 & 4\\ \hline 
         X7& 3 & 4 & 5\\ \hline 
         X8& 3 & 4 & 4\\ \hline 
         X9& 3 & 4 & 5\\ \hline 
         X10& 2 & 4 & 4\\ \hline 
         X11& 1 & 3 & 4\\ \hline 
         X12& 2 & 3 & 4\\ \hline 
         X13& 1 & 2 & 3\\ \hline
    \end{tabular}
    \caption{Prospective scenarios considered in our analysis}
    \label{tab:scenarios}
\end{table}

The \textbf{pessimistic scenario} is one where attributes related to i) participants, ii) TTX execution, and iii) presented scenario, are valued with a low or moderate score, having as consequence a low satisfaction ($Y=2.8$), and a low ability to apply the knowledge acquired along the TTX for the resolution of a CSA situation ($X13=1$). The lack of experience and preparation in key areas results in participants struggling to make informed decisions under crisis conditions.
Regarding attributes related to participants, this scenario considers situations where participants have scarce experience in general TTXs, and almost null experience or knowledge in cybersecurity simulations, maritime cyberdefense exercises, and cybersecurity and cyberdefense concepts, i.e. ${X1=1,X2=0,X3=0,X4=0}$. Regarding attributes related to TTX execution, this scenario assumes situations where the facilitator has a moderate performance in terms of instructions, time management and delivered resources, i.e. ${X5=3,X6=3,X7=3}$. Finally, regarding attributes related to the presented scenario, this scenario considers cases where the studied case has deficiencies in terms of relevance, easy-to-understand, complexity, and contributed-details, i.e. ${X8=3,X9=3,X10=2,X11=1,X12=2}$.

The \textbf{trend-based scenario} is one where attributes related to i) participants, ii) TTX execution, and iii) the presented scenario are valued with medium scores, resulting in a moderate level of satisfaction ($Y=3.8$) and a medium-low ability to apply the knowledge acquired during the TTX for resolving a CSA situation ($X13=2$). This outcome reflects incremental improvements compared to the pessimistic scenario, particularly in aspects related to TTX execution and the scenario itself. While participants benefit from some prior training in cybersecurity and cyberdefense, and the TTX execution is better structured, the scenario complexity remains moderate, limiting the depth of decision-making improvement. These improvements lead to more informed decision-making, but gaps still exist in the application of acquired knowledge during a cyber crisis.
The trend-based scenario assumes that participants have some prior experience in crisis simulation exercises or TTX ${X1=1}$, but limited experience in cybersecurity and cyberdefense simulations ${X2=0}$ and maritime-specific cyberdefense simulations ${X3=0}$. Additionally, participants have received some form of formal or informal training in cybersecurity and cyberdefense ${X4=1}$, which provides a foundational understanding of key concepts but lacks practical application in high-pressure scenarios. In terms of TTX execution, the facilitator performs well, providing clear instructions, managing time effectively, and delivering adequate resources, i.e. ${X5=4,X6=4,X7=4}$. However, the facilitator’s performance, while improving compared to the pessimistic scenario. Finally, regarding the presented scenario, the case exhibits moderate relevance and is relatively easy to understand, but the complexity remains at an intermediate level, meaning the scenario presents a reasonable challenge but does not push participants to their maximum decision-making limits. The scenario also provides a moderate level of detail, enough to foster learning but not sufficient for comprehensive, advanced-level strategic decision-making, i.e. ${X8=4,X9=4,X10=4,X11=3,X12=3}$.

The \textbf{optimistic scenario} is one where attributes related to i) participants, ii) TTX execution, and iii) the presented scenario are valued with high scores, resulting in a high level of satisfaction ($Y=4.8$) and a high ability to apply the knowledge acquired during the TTX for resolving a CSA situation ($X13=3$). This scenario reflects the best-case scenario, where all key aspects align to maximize learning and decision-making capabilities. Participants are well-prepared, the TTX execution is highly effective, and the presented scenario offers the highest level of relevance and challenge. These conditions allow participants to make the most informed decisions, confidently applying their knowledge in resolving complex cyber crises.
The optimistic scenario assumes that participants have significant prior experience in crisis simulation exercises or TTX ${X1=1}$, and have participated in cybersecurity and cyberdefense simulations ${X2=1}$, including maritime-specific cyberdefense simulations ${X3=1}$. Additionally, participants have received formal or informal training in cybersecurity and cyberdefense ${X4=1}$, which provides them with advanced knowledge and practical skills. In terms of TTX execution, the facilitator performs excellently, providing clear instructions, managing time effectively, and delivering optimal resources ${X5=5, X6=4, X7=5}$. The facilitator’s role in providing dynamic, real-time feedback during critical phases ensures that participants are fully engaged and able to learn from the simulation. Finally, regarding the presented scenario, the case exhibits high relevance, is easy to understand, and presents high complexity, challenging participants to apply advanced decision-making skills under pressure. The scenario provides a high level of detail, allowing participants to develop comprehensive, strategic responses to complex cyber crises ${X8=5, X9=4, X10=4, X11=4, X12=4}$.

The prospective scenario analysis indicates that the success of TTX exercises and cybersecurity simulations depends significantly on their design and relevance to participants. Strengthening participation in maritime cybersecurity and cyberdefense simulations ($X3$) is crucial, as the difference in satisfaction between the pessimistic and optimistic scenarios suggests that prior simulation experience ($X1$) is a key factor. Ensuring well-trained facilitators with dynamic methodologies can improve participant engagement ($X5$), while realistic and well-structured scenarios enhance immersion ($X8 - X12$). Effective time management ($X6$) and resource ($X7$) allocation are also essential elements that influence participant overall satisfaction ($Y$), requiring careful planning and execution. Strategic adjustments in these areas can significantly improve the perception and effectiveness of future exercises.

\section{Conclusions}\label{conclusions}

This study introduces a novel hybrid TTX methodology for maritime cybersecurity and cyberdefense, integrating traditional TTX techniques with a mathematical simulation-based model (SERDUX-MARCIM). This approach creates a dynamic and realistic training environment, allowing participants to engage with simulated cyber threats and make informed decisions based on real-time data. By combining serious gaming with simulation modeling, it addresses key limitations of traditional TTX, enhancing both the learning experience and decision-making capabilities.

The methodology was successfully applied in two exercises in Buenos Aires and Washington D.C., using real-world cyber crisis scenarios that demonstrated its effectiveness in strengthening CSA among strategic-level stakeholders in the maritime sector, including military personnel, maritime authorities, and policymakers. These exercises provided valuable insights into improving crisis management and response capabilities. The positive feedback from participants highlights the methodology’s potential for broad adoption in maritime cybersecurity.

For future work, it is proposed to increase both the number of participants and the number of exercises conducted, thereby enhancing the statistical robustness and reliability of the results. With larger sample sizes, dimensionality reduction techniques could also be considered to optimize the regression model and focus on the most significant variables. Additionally, the implementation of a pre-exercise assessment is recommended to measure participants' progress throughout the activity. Finally, aligning the evaluated competencies with recognized cybersecurity education and training standards will strengthen the methodological rigor and educational relevance of the proposed TTX.

\bibliographystyle{IEEEtran}
\bibliography{bibliography.bib}{}

\end{document}